\documentstyle[prd,aps,twocolumn,epsfig]{revtex}
\newcommand{\be}{\begin{eqnarray}}
\newcommand{\ee}{\end{eqnarray}}

\def\lsim{\mathrel{\rlap{\lower4pt\hbox{\hskip1pt$\sim$}}
    \raise1pt\hbox{$<$}}}               % less than or approx. symbol
\def\gsim{\mathrel{\rlap{\lower4pt\hbox{\hskip1pt$\sim$}}
    \raise1pt\hbox{$>$}}}               % greater than or approx. symbol

\begin{document}

\title{
\rightline{\large{Preprint RM3-TH/00-2}}
\vspace{0.5cm}
Parton picture of inclusive quasi-elastic electron scattering off nuclei
}

\author{Silvano Simula}
\address{Istituto Nazionale di Fisica Nucleare, Sezione Roma III, Via della Vasca Navale 84, I-00146 Roma, Italy}

\maketitle

\begin{abstract} 

\noindent A parton picture of inclusive quasi-elastic electron-nucleus scattering, where individual nucleons are treated as (non point-like) partons of the nucleus, is presented. All the necessary target-mass corrections to asymptotic scaling are evaluated and a new scaling variable $\xi_{QE}$ is obtained. Recent inclusive data from Jefferson Lab are analyzed and shown to scale in terms of $\xi_{QE}$. The new approach allows to extract in a model-independent way the nucleon light-cone momentum distribution in the nucleus.


\end{abstract}

\section{Introduction}

\indent Inclusive quasi-elastic ($QE$) electron-nucleus scattering is widely recognized as an important tool for extracting unique information on the dynamics of nucleons in nuclei, in particular on the nature of short-range  and tensor nucleon-nucleon ($NN$) correlations \cite{Strikman}. In the $QE$ kinematical regions the dominant reaction mechanism is expected to be the elastic scattering from individual nucleons bound in the nucleus. Therefore, the concept of $y$-scaling was introduced in \cite{West}, showing that at sufficiently high values of the squared four-momentum transfer $Q^2$ ($= - q \cdot q = \nu^2 - |\vec{q}|^2$) the measured $QE$ cross section is predicted to scale in terms of the variable $y$, which represents the minimum momentum of the struck nucleon along the direction $\vec{q}$ of the virtual photon. Adopting the non-relativistic approximation for the nuclear wave function, the scaling function $F(y)$ was shown \cite{West} to be the nucleon longitudinal momentum distribution, $f(y)$, which depends only on the ground-state of the target nucleus.

\indent Existing inclusive $SLAC$ measurements \cite{SLAC} do not exhibit scaling at $y > 0$, because of the relevance of competitive mechanisms, like meson-exchange currents at low $Q^2$ and nucleon inelastic channels at high $Q^2$. However, in the low-energy side of the $QE$ peak (corresponding to $y < 0$) the $QE$ mechanism is dominant and the $SLAC$ data show indeed an approximate scaling for low negative values of $y$. At $y \lsim - 0.3 ~ GeV/c$, corresponding to the kinematical regions relevant for the investigation of $NN$ correlations, the $y$-scaling breaks down because of the effects from the final state interactions ($FSI$) among the struck nucleon and the residual nuclear system (cf. \cite{FSI}). Thus, the standard approach \cite{Ciofi} is to extract the scaling function $F(y)$ by extrapolation from finite $Q^2$ to the limit $Q^2 \to \infty$, where the $FSI$ are expected to vanish. However, a careful theoretical analysis has shown \cite{Ciofi} that $F(y) \neq f(y)$ because of the effects of the nucleon removal energy distribution in the nucleus (binding effects). Since microscopic calculations of the latter are still lacking (except for very light nuclei) and only models are available (cf. \cite{PKE}), the y-scaling does not allow to extract the nucleon longitudinal momentum distribution in a model-independent way \cite{note}.

\indent The aim of this work is to present concisely a new approach to scaling for inclusive $QE$ electron-nucleus scattering, in which the individual nucleons are treated as the (non point-like) partons of the nucleus. All the target-mass effects, due to the non-vanishing masses of the struck particle (the nucleon mass $M$) and of the target system (the nuclear mass $M_A \simeq A M$), are evaluated and a {\em new scaling variable} $\xi_{QE}$ is obtained, viz.
 \be
    \xi_{QE} \equiv x {1 + \sqrt{1 + 4M^2/Q^2} \over 1 + \sqrt{1 + 
    4M^2x^2/Q^2}}
    \label{eq:xiQE}
 \ee
where $x = Q^2 / 2 M \nu$ is the Bjorken variable ($0 \leq x \leq M_A /M \simeq A$). Note that the low-energy side of the $QE$ peak ($x \geq 1$) corresponds to $\xi_{QE} \geq 1$ with $\xi_{QE} \leq x$. Moreover, $\mbox{lim}_{Q^2 \to \infty} ~ \xi_{QE} = x$. The variable $\xi_{QE}$ has the physical meaning of the light-cone fraction of the nucleus momentum carried by the struck nucleon and the resulting scaling function is proportional to the nucleon light-cone momentum distribution in the nucleus, which depends only on the target ground-state (cf., e.g., \cite{EMC}). The concept of $\xi_{QE}$ scaling is then positively checked against the recent inclusive data from $JLab$ \cite{JLAB}, which extend the $Q^2$ range of the previous $SLAC$ measurements up to $Q^2 \sim 7 ~ (GeV/c)^2$. Finally, the nucleon light-cone momentum distribution in $^{56}Fe$ is extracted from the $JLab$ data in a model-independent way.

\section{Parton picture of $QE$ scattering}

\indent As it is well known, the parton model \cite{parton} successfully describes the deep inelastic electron-nucleon scattering process in terms of elastic processes on free (point-like) constituents, the partons. At sufficiently large values of $Q^2$ the nucleon structure function is predicted to scale in the Bjorken variable $x$ and to be directly related to the parton light-cone momentum distribution in the nucleon $\rho_q(x)$ as
\be
     \label{eq:F2N}
     F_2^N(x, Q^2) \to_{Bj} x \sum_q e_q^2 ~ \rho_q(x)
     \ee
where $e_q$ is the charge of the parton $q$. The predictions of the parton model are fully confirmed by the analysis of the light-cone singularities of the time-ordered product of quark currents. Moreover, using the Operator Product Expansion ($OPE$) formalism \cite{Wilson} it has been shown  \cite{GP76} that an approximate scaling can still hold {\em at finite values of $Q^2$}, provided target-mass corrections are properly taken into account. To this end a new scaling variable $\xi \neq x$, known as the Nachtmann variable \cite{NAC73}, has to be introduced, depending on $Q^2$ and the target mass (the nucleon mass), viz.
 \be
      \label{eq:csi}
       \xi = {2x \over 1 + \sqrt{1 + 4M^2x^2/ Q^2}}.
 \ee

\indent Violations of the $\xi$ scaling are related to dynamical higher-twist contributions, which manifest themselves as power suppressed terms (i.e., powers of $1 / Q^2$), and to logarithmic $pQCD$ corrections (cf. \cite{GP76}).

\indent Let us now consider in a similar picture the $QE$ scattering off nuclei, being inspired by the fact that the $QE$ mechanism is an {\em elastic} process on the individual nucleons in the nucleus. Thus, at sufficiently large values of $Q^2$ we expect that the $QE$ contribution to the nuclear structure function (which will be simply denoted hereafter by $F_2^A$) can be written in terms of the nuclear Bjorken variable $x_A = Q^2 / 2 M_A \nu$ as
 \be
       F_2^A(x_A, Q^2) \to x_A \sum_{N=1}^A ~ G_N^2(Q^2) ~ \rho_N(x_A)
      \label{eq:F2A}
 \ee
where $0 \leq x_A \leq 1$, $\rho_N(x_A)$ is the nucleon light-cone momentum distribution in the nucleus (appearing also in the nucleonic contribution to the $EMC$ effect \cite{EMC}), while $G_N^2(Q^2)$ generalizes the squared parton charge $e_q^2$ to the case of nucleons, viz.
 \be
      \label{eq:GN}
      G_N^2(Q^2) \equiv {G_E^2(Q^2) + \tau G_M^2(Q^2) \over 1 + \tau}
 \ee
with $\tau \equiv Q^2 / 4M^2$. For point-like nucleons one has $G_E(Q^2), ~ G_M(Q^2) \to e_N$, implying $G_N^2(Q^2) \to e_N^2$. Note that the distribution $\rho_N(x_A)$, appearing in Eq. (\ref{eq:F2A}), satisfies naturally both the baryon ($\sum_{N=1}^A \int_0^1 dx_A \rho_N(x_A) = A$) and momentum sum rules ($\sum_{N=1}^A \int_0^1 dx_A x_A \rho_N(x_A) = 1$), provided the nucleons are the only relevant degrees of freedom in the nucleus (cf. \cite{EMC}). Let us point out that Eq. (\ref{eq:F2A}) does not exhibit a scaling property because of the $Q^2$-dependence of the nucleon form factors. Therefore, in what follows we will consider {\em reduced} nuclear structure functions, defined as
 \be
     \label{eq:F2A_reduced}
     \hat{F}_2^A(x_A, Q^2) \equiv  {F_2^A(x_A, Q^2) \over \sum_{N=1}^A 
     G_N^2(Q^2)}.
 \ee
Basically, in the kinematical regions of interest in this work the $Q^2$-dependence of $\hat{F}_2^A(x_A, Q^2)$ drops out. Indeed, assuming the dipole-law for $G_E^N(Q^2)$ and $G_M^N(Q^2)$, one gets
 \be
      \label{eq:F2A_hat}
      \hat{F}_2^A(x_A, Q^2) \simeq  x_A \sum_{j = n, p} \tilde{e}_j^2(Q^2) ~  
       \rho_j(x_A)
 \ee
where $\tilde{e}_p^2(Q^2) = (1 + \tau \mu_p^2) / [1 + \tau (\mu_p^2 + \mu_n^2 N / Z)]$ and $\tilde{e}_n^2(Q^2) = 1 - \tilde{e}_p^2(Q^2)$ play the role of squared effective charges for protons and neutrons, respectively. For $Q^2 \gsim 1 ~ (GeV/c)^2$one has $\tilde{e}_p^2(Q^2) \simeq 1 / (1 + 0.47 N / Z)$.

\indent We have now to find the variable that generalizes the  Nachtmann variable $\xi$ to the case of $QE$ scattering, where both the mass of the struck particle (the nucleon mass $M$) and the mass of the target system (the nuclear mass $M_A \simeq A M$) have to be considered. (For an investigation of nuclear scaling in terms of the variable $\xi$ see \cite{Filippone}). Let us consider first the effects of the target mass (the nuclear mass $M_A$), which lead to the following nuclear Nachtmann variable: $\xi_A = 2 x_A / [1+ \sqrt{1 + 4M_A^2x_A^2 / Q^2}]$. However, since $M_A x_A = M x$, the variable $\xi_A$ turns out to be simply a rescaled Nachtmann variable, i.e. $\xi_A = \xi M / M_A \simeq \xi / A$ where now $0 \leq \xi \leq A$. The effects of the struck mass (the nucleon mass $M$) can be easily included in our partonic picture. Indeed, since partons are assumed to be free (and therefore on-mass-shell), one gets the conditions $(k_N + q)^2 = M^2$ and $k_N^2 = M^2$. In terms of light-cone variables we can put $k_N = (k^+, M^2 / k^+, \vec{0}_{\perp})$ and $q = (-M_A \xi_A, Q^2 / M_A \xi_A, \vec{0}_{\perp})$, where we have used the relation $|\vec{q}| - \nu =  M \xi = M_A \xi_A$. Thus, one gets the polynomial equation $k^{+2} - M_A \xi_A k^+ - M^2 M_A^2\xi_A^2 / Q^2 = 0$, which implies $k^+ = (M_A \xi_A / 2) [1 + \sqrt{1 + 4M^2 / Q^2}]$. Introducing the (rescaled) light-cone fraction $z = k^+ / M$ ($0 \leq z \leq A$) we obtain $z = \xi_{QE}$ with $\xi_{QE}$ given by Eq. (\ref{eq:xiQE}), representing therefore the appropriate variable in which we expect to observe approximate scaling for the $QE$ (reduced) nuclear structure function {\em at finite values of $Q^2$}. We have to mention that a scaling variable conceptually similar to $\xi_{QE}$ was introduced in \cite{Strikman}(b).

\indent Since the variable $\xi_{QE}$ allows to take into account kinematical (target-mass) corrections,  violations of the $\xi_{QE}$ scaling are expected to be related to dynamical effects, like those due to $FSI$, which should appear as power suppressed terms, i.e. powers of $1 / Q^2$ (see next Section). Finally note that the scaling variable $\xi_{QE}$ does not depend on the mass number $A$ and therefore it is the same for all nuclei.

\indent We now derive the target-mass corrections to the nuclear $QE$ structure functions. Let us first consider the case of point-like nucleons and then insert the effects of the nucleon size where appropriate. For point-like nucleons we can adapt the procedure of \cite{Barbieri}, where the case of scattering from massive quarks in the nucleon was considered. Let us start from the nuclear forward Compton amplitude (cf. \cite{Ji})
\be
     \label{eq:Compton}
     T_{\mu \nu}^A(P, q) & = & {i \over \pi} \int d^4z e^{iqz} \langle P| 
     {\cal{T}}[J_{\mu}(z) J_{\nu}(0)] |P \rangle = \nonumber \\
     & - & T_1^A(\nu, Q^2) \left( g_{\mu \nu} + {q_{\mu} q_{\nu} \over Q^2} 
     \right) \nonumber \\
     & + & {1 \over M_A^2} T_2^A(\nu, Q^2) \tilde{P}_{\mu} \tilde{P}_{\nu}
 \ee
where $|P \rangle$ is the nuclear ground-state with total four-momentum $P$, $\tilde{P}_{\mu} = P_{\mu} + (q \cdot P) ~ q_{\mu} / Q^2$ and $J_{\mu}(z)$ is the nuclear current given by $J_{\mu}(z) = \bar{\psi}(z) \gamma_{\mu} \psi(z)$, with $\psi(z)$ being the nucleon field. The singularities generated in the product $J_{\mu}(z) J_{\nu}(0)$ as $z \to 0$ can be treated within the framework of the $OPE$ and it has been shown \cite{NAC73,Wandzura} that such an $OPE$ can be rewritten as an expansion over Gegenbauer polynomials, viz.
 \be
      \label{eq:Gegenbauer_1}
      3T_1^A - (1 + {\nu^2 \over Q^2}) T_2^A & = & {2 \over \pi} \sum_{n \ge 
      2} C_n^1(i {\nu \over Q}) sin[{\pi \over 2} (n + 1)] \nonumber \\
      & \cdot & \left( {M_A \over Q} \right)^n \mu_n^{(1)}(Q^2)
 \ee
and
 \be
      \label{eq:Gegenbauer_2}
      T_2^A & = & - {8 \over \pi} \sum_{n \ge 2} C_{n-2}^3(i {\nu \over Q}) 
      sin[{\pi \over 2} (n + 1)] \nonumber \\
      & \cdot &\left( {M_A \over Q} \right)^n \mu_n^{(2)}(Q^2)
 \ee
where $C_n^{\lambda}$ are Gegenbauer polynomials and $\mu_n^{(i)}(Q^2)$ are Nachtmann moments ($i = 1, 2$). Using standard dispersion relations for $T_i^A(\nu, Q^2)$ (cf. the nucleon case in \cite{NAC73,Wandzura}) and orthogonality relations among Gegenbauer polynomials, from Eqs. (\ref{eq:Gegenbauer_1}-\ref{eq:Gegenbauer_2}) it follows
 \be
       \label{eq:mu1}
       \mu_n^{(1)} = {2M_A \over Q^2} \int d\nu \xi_A^{n+1} \left[ 3W_1^A - 
       {\nu^2 + Q^2  \over Q^2} W_2^A \right]
 \ee
and
 \be
       \label{eq:mu2}
       \mu_n^{(2)} & = & {4M_A \over Q^4} \int d\nu \xi_A^{n+1} W_2^A 
       \left[ (n^2 + 2n + 3) \nu^2 \right. \nonumber \\
       & + & \left. n(n + 2) Q^2 + 3(n + 1) \nu \sqrt{\nu^2 +  Q^2} \right] 
       \nonumber \\
       & /  & \left[ (n + 2) (n + 3) \right]
 \ee
where $W_i^A(\nu, Q^2)$ are the structure functions of the absorptive part of $T_{\mu \nu}^A$ (i.e., of the nuclear tensor). As noticed in \cite{Barbieri}, the Nachtmann moments $\mu_n^{(i)}(Q^2)$ can be written as $\mu_n^{(i)}(Q^2) = \tilde{\mu}_n^{(i)}(Q^2) \cdot \langle P | \hat{O}_n | P \rangle$, where $\langle P| \hat{O}_n | P \rangle$ are the reduced matrix elements of the traceless operators appearing in the $OPE$ of the nuclear tensor and yielding the scaling function, while $\tilde{\mu}_n^{(i)}(Q^2)$ are the Nachtmann moments corresponding to the sum of the individual parton contributions in $W_i^A(\nu, Q^2)$, namely: $W_1^A \to \tau \sum_{N=1}^A e_N^2 \delta(\nu - Q^2 / 2M)$ and $W_2^A \to \sum_{N=1}^A e_N^2 \delta(\nu - Q^2 / 2M)$. The resummation of the series over the Gegenbauer polynomials in Eqs. (\ref{eq:Gegenbauer_1}-\ref{eq:Gegenbauer_2}) can be done in the same way as in \cite{GP76,Barbieri} and the final result for the nuclear structure function $\nu W_2^A(\xi_{QE}, Q^2)$ reads as
 \be
       \nu W_2^A = {x^2 \over r^3} \left[ (1 + {1 \over \tau}) 
       {F_2^A(\xi_{QE}) \over \xi_{QE}^2} + 3 T_2^A(\xi_{QE}, Q^2)  \right]
       \label{eq:nuW2A}
 \ee
where $r \equiv \sqrt{1 + 4M^2x^2 / Q^2}$ and
 \be
      T_2^A & = & \sqrt{1 + {1 \over \tau}} {2M^2 x \over Q^2 r}  
      \int_{\xi_{QE}}^{\xi_{QE}^{max}} d\xi_{QE}^{\prime}  \left(1 - {1 
      \over \xi_{QE}^{\prime 2}} \right) 
      \nonumber \\
      & \cdot & {F_2^A(\xi_{QE}^{\prime}) \over \xi_{QE}^{\prime 2}} + {4M^4 
      x^2 \over Q^4 r} \int_{\xi_{QE}}^{\xi_{QE}^{max}} d\xi_{QE}^{\prime} 
      \left( 1 - {1 \over \xi_{QE}^{\prime 2}} \right) 
      \nonumber \\
      & \cdot & {F_2^A(\xi_{QE}^{\prime}) \over \xi_{QE}^{\prime 2}} \left[ 
      \xi_{QE}^{\prime} + {1 \over \xi_{QE}^{\prime}} - \xi_{QE} - {1 \over 
      \xi_{QE}} \right]
      \label{eq:T2A}
 \ee
with $\xi_{QE}^{max} = \mbox{min}[A, Q (1+ \sqrt{1 + 4M^2 / Q^2}) / 2M]$ (cf. \cite{SIM00}). The function $F_2^A(\xi_{QE})$ is the asymptotic limit of $\nu W_2^A(\xi_{QE}, Q^2)$, i.e. at fixed $\xi_{QE}$ one has
 \be
      \nu W_2^A  \to_{Q^2 \to \infty}  F_2^A(\xi_{QE}) = \xi_{QE} \sum_{N 
      = 1}^A e_N^2 ~ \rho_N(\xi_{QE}).
      \label{eq:asymptotic}
 \ee 

\indent As for the nuclear response $W_1^A(\nu, Q^2)$,  one can note that for the combination $\{3W_1^A -  W_2^A [\nu^2 + Q^2]  / Q^2\}$ Eq. (\ref{eq:mu1}) is similar to the Nachtmann moment of a scalar current (cf. \cite{NAC73,Barbieri}). Thus, the target-mass-corrected nuclear response $W_1^A(\xi_{QE}, Q^2)$ is given by
 \be
      W_1^A = {x \over 2Mr} \left[ {1 + 1 / \tau \over 1 + R_N} 
      {F_2^A(\xi_{QE}) \over \xi_{QE}^2} +  T_2^A(\xi_{QE}, Q^2) \right]
      \label{eq:W1A}
 \ee
where $R_N(Q^2) = 1 / \tau$ is the longitudinal to transverse ($L / T$) cross section ratio for point-like nucleons.

\indent The modifications of our basic equations (\ref{eq:nuW2A}-\ref{eq:W1A}) due to the nucleon size can be argued in the following way. From $y$-scaling analysis it is known (cf., e.g., \cite{FSI,Ciofi}) that the quantity $\sum_{N=1}^A G_N^2(Q^2)$ can be factorized out, so that the scaling function $F(y)$ describes the asymptotic nuclear response as the nucleons were point-like in the virtual photon coupling. Therefore, we expect that  Eqs. (\ref{eq:nuW2A}-\ref{eq:asymptotic}) still hold by replacing $\nu W_2^A$, $F_2^A$ and $T_2^A$ with their reduced counterparts, defined as in Eq. (\ref{eq:F2A_reduced}). Correspondingly, in Eq. (\ref{eq:W1A}) we expect the replacements of $W_1^A$, $F_2^A$ and $T_2^A$ with their reduced counterparts and, furthermore, the replacement of $L /T$ ratio $R_N(Q^2) = 1 / \tau$ with
 \be
      \label{eq:RN}
      \hat{R}_N(Q^2) \equiv {1 \over \tau} {\sum_{N^{\prime} =1}^A 
      [G_E^{N^{\prime}}(Q^2)]^2  \over \sum_{N^{\prime} =1}^A 
      [G_M^{N^{\prime}}(Q^2)]^2}.
 \ee 

\indent It is interesting to note that the above expectations may follow from the {\em hypothesis} that the Gegenbauer expansions (\ref{eq:Gegenbauer_1}-\ref{eq:Gegenbauer_2}) hold as well in case of non point-like nucleons. Indeed, within this hypothesis, in constructing the moments $\tilde{\mu}_n^{(i)}(Q^2)$ one has to consider in $W_i^A(\nu, Q^2)$ the sum of the individual nucleon contributions given now by $W_1^A \to \tau \sum_{N=1}^A [G_M^N(Q^2)]^2 \delta(\nu - Q^2 / 2M)$ and $W_2^A \to \sum_{N=1}^A G_N^2(Q^2) \delta(\nu - Q^2 / 2M)$. It is then easy to check that in Eqs. (\ref{eq:mu1}-\ref{eq:mu2}) the quantity $\sum_{N=1}^A G_N^2(Q^2)$ does factorize out and that the $L / T$ ratio $R_N(Q^2) = 1 / \tau$ has to be replaced by $\hat{R}_N(Q^2)$ given in Eq. (\ref{eq:RN}). In the next Section we will show that our basic equations and the $\xi_{QE}$ variable are successful in describing the scaling in the $QE$ process, providing therefore a challenge to demonstrate that Eqs. (\ref{eq:Gegenbauer_1}-\ref{eq:Gegenbauer_2}) hold as well for non point-like nucleons. To this end a possibility (which can be investigated in future works) might be offered by the application of nuclear effective field theories.

\indent Note that according to Eq. (\ref{eq:nuW2A}) the structure function $\nu W_2^A(\xi_{QE}, Q^2)$ [$\nu \hat{W}_2^A(\xi_{QE}, Q^2)$] does not depend on the nucleon $L / T $ ratio $R_N(Q^2)$ [$\hat{R}_N(Q^2)$]. This is at variance with the results of the impulse approximation obtained within the instant form of the dynamics (cf., e.g., \cite{SIM95}), while it agrees with the results of the light-cone impulse approximation (cf. \cite{EMC}). This happens because our results are based on a partonic description which is naturally formulated in the light-cone form of the dynamics. As a matter of fact, the impulse approximations in the instant and the light-cone forms should coincide only asymptotically, while they can differ at finite values of $Q^2$.

\indent A very relevant feature of the target-mass corrections is that the asymptotic function $\hat{F}_2^A(\xi_{QE})$ can be extracted from the structure functions $\hat{W}_i^A(\xi_{QE}, Q^2)$ {\em at finite values of $Q^2$}, viz.
 \be
       \hat{F}_2^A(\xi_{QE}) & = & {r^3 \xi_{QE}^2 \over x^2 (1 + 1 / \tau)} 
       {1 + \hat{R}_N(Q^2) \over 2 - \hat{R}_N(Q^2)} \nonumber \\
       & \cdot & {2 - R_A(\xi_{QE}, Q^2) \over 1 + R_A(\xi_{QE}, Q^2)} \nu 
       \hat{W}_2^A(\xi_{QE}, Q^2)
       \label{eq:F2A_TM}
 \ee
where
 \be
      \label{eq:RA}
      R_A(\xi_{QE}, Q^2) \equiv {W_2^A(\nu, Q^2) \over W_1^A(\nu, Q^2)} 
      \left( 1 + {\nu^2 \over Q^2} \right) - 1
 \ee
is the nuclear $QE$ longitudinal to transverse cross section ratio. In other words Eq. (\ref{eq:F2A_TM}) tells us how to cancel out all the target-mass corrections from the structure function $\nu \hat{W}_2^A(\xi_{QE}, Q^2)$.

\section{Results}

\indent Let us now consider the {\em experimental} nuclear structure function  $\nu \bar{W}_2^A(\xi_{QE}, Q^2)$, which in terms of the inclusive nuclear cross section $\sigma_A$ is given by
 \be
      \nu \bar{W}_2^A & = & {\nu \sigma_A \over \sigma_{Mott} \cdot [\sum_{N 
      = 1}^A G_N^2(Q^2)]} \nonumber \\
      & \cdot & {1 \over 1 + 2 tg^2({\theta_e \over 2}) (1 + \nu^2 / Q^2) / 
      (1+ R_A)}
      \label{eq:nuW2A_exp}
 \ee
where $\sigma_{Mott}$ is the Mott cross section and $\theta_e$ is the electron scattering angle. We do not expect to observe scaling of Eq. (\ref{eq:nuW2A_exp}) in terms of $\xi_{QE}$, at least because of the $Q^2$-dependence induced by target-mass corrections (see Eq. (\ref{eq:nuW2A})). Significative scaling violations are indeed present in the iron data of \cite{JLAB} at any $\xi_{QE}$, as it is illustrated in Fig. 1, where the reported uncertainties include the small impact of the variation of $R_A$ in Eq. (\ref{eq:nuW2A_exp}) between $0$ and $2\hat{R}_N$.

\indent Inspired by Eq. (\ref{eq:F2A_TM}) let us define the following {\em experimental} scaling function:
 \be
       \bar{F}_2^A(\xi_{QE}, Q^2) & \equiv & {r^3 \xi_{QE}^2 \over x^2 (1 + 
       1 / \tau)} {1 +  \hat{R}_N(Q^2) \over 2 - \hat{R}_N(Q^2)} 
       \nonumber \\
       & \cdot & {2 - R_A(\xi_{QE}, Q^2) \over 1 + R_A(\xi_{QE}, Q^2)} 
       \nu \bar{W}_2^A(\xi_{QE}, Q^2).
       \label{eq:F2A_exp}
 \ee

\noindent Therefore, from the results of the previous Section we expect that
 \be
       \bar{F}_2^A(\xi_{QE}, Q^2) = \xi_{QE} ~ \bar{\rho}(\xi_{QE}) + O({1 
       \over Q^2})
       \label{eq:HT}
 \ee
where $\bar{\rho}(\xi_{QE})$ is the nucleon light-cone momentum distribution averaged over protons and neutrons in the nucleus [cf. Eq. (\ref{eq:F2A_hat})], while the power-suppressed terms correspond to higher twists containing unique information about the structure of $FSI$ and/or other relevant mechanisms which violate the $\xi_{QE}$ scaling. Note that Eq. (\ref{eq:HT}) implies that any $Q^2$-dependence of $\bar{F}_2^A(\xi_{QE}, Q^2)$ signals {\em the breakdown of the impulse approximation}.

\begin{figure}[htb]

\vspace{0.25cm}

\centerline{\epsfxsize=7.50cm \epsfig{file=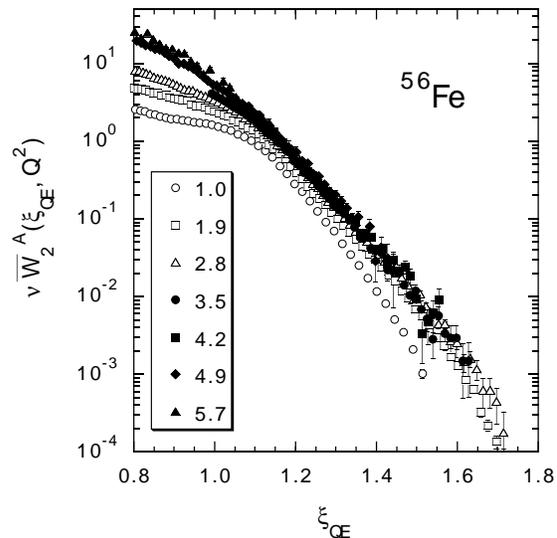}}

\caption{Results of Eq. (\ref{eq:nuW2A_exp}) evaluated for the iron data of \protect\cite{JLAB} versus $\xi_{QE}$ (Eq. (\ref{eq:xiQE})). The inset shows the value of $Q^2$ corresponding to the top of the $QE$ peak ($\xi_{QE} = 1$).}

\vspace{0.25cm}

\end{figure}

\indent The results of Eq. (\ref{eq:F2A_exp}) evaluated for the iron data of \cite{JLAB} are shown in Fig. 2. It can clearly be seen that scaling violations are present for $\xi_{QE} < 1.1$ because of the large contributions from nucleon inelastic channels (cf. \cite{JLAB}), but now the data free from target-mass effects scale nicely for $1.1 \lsim \xi_{QE} \lsim 1.4$, while for larger $\xi_{QE}$ the scaling is partially broken by $FSI$ effects.

\indent Applying a power correction analysis to the data of Fig. 2 (after the subtraction of the nucleon inelastic channel contribution evaluated as in \cite{SIM95}), the asymptotic function $\hat{F}_2^A(\xi_{QE}) = \xi_{QE} ~ \bar{\rho}(\xi_{QE})$ can be extracted in a model-independent way and the resulting nucleon light-cone momentum distribution $\bar{\rho}(\xi_{QE})$ is shown in Fig. 3. We point out that our procedure is fully relativistic and does not involve the non-relativistic approximation for the nuclear wave function; moreover, being formulated in the light-cone form of the dynamics, it does not suffer from any ambiguities related to the modelling of binding effects and to off-shell prescriptions.

\begin{figure}[htb]

\vspace{0.25cm}

\centerline{\epsfxsize=7.50cm \epsfig{file=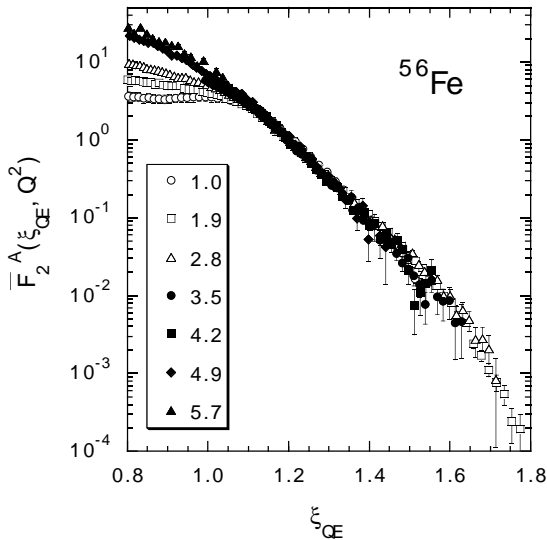}}

\caption{Results of Eq. (\ref{eq:F2A_exp}) evaluated for the iron data of \protect\cite{JLAB} versus $\xi_{QE}$ (Eq. (\ref{eq:xiQE})). The inset is as in Fig. 1.}

\vspace{0.25cm}

\end{figure}

\begin{figure}[htb]

\vspace{0.25cm}

\centerline{\epsfxsize=7.50cm \epsfig{file=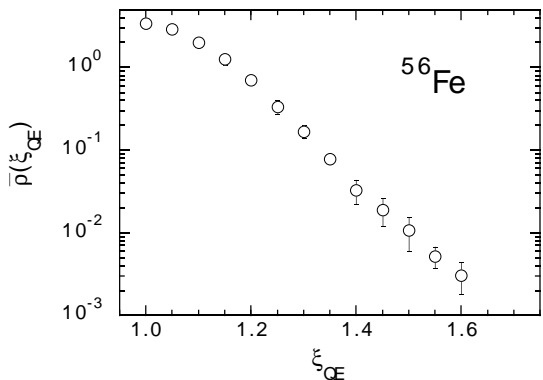}}

\caption{The nucleon light-cone momentum distribution $\bar{\rho}(\xi_{QE})$  (see Eq. (\ref{eq:HT})), extracted from a power correction analysis of the iron data of Fig. 2, versus $\xi_{QE}$ (Eq. (\ref{eq:xiQE})).}

\vspace{0.25cm}

\end{figure}

\indent Finally, since the variable $\xi_{QE}$ and the Nachtmann variable $\xi$ are simply related by $\xi_{QE} = \xi ~ [1 + \sqrt{1 + 4M^2 / Q^2}]/2$, one could use $\xi$ as a scaling variable for the $QE$ process (cf. \cite{Filippone}). From the theoretical point of view it is however clear that the variable $\xi$ is appropriate only for elastic processes on massless constituents in a massive target. Consequently, all the machinery of $QE$ target-mass corrections developed in this work cannot be formulated in terms of $\xi$. Therefore, any approximate phenomenological scaling in $\xi$ should be simply reminiscent of the scaling in $\xi_{QE}$, which we stress is the appropriate variable for the $QE$ process because it includes the mass effects of both the struck particle and the target.

\section{Conclusions}

\indent In conclusion, a new approach to scaling in inclusive quasi-elastic electron-nucleus scattering has been concisely presented. It is based on a parton picture of the quasi-elastic process, where individual nucleons are treated as the (non point-like) partons of the nucleus. All the target-mass corrections to asymptotic scaling have been taken into account, leading to a {\em new scaling variable} $\xi_{QE}$ given by Eq. (\ref{eq:xiQE}). The $\xi_{QE}$ scaling has been positively checked against recent inclusive iron data from Jefferson Lab \cite{JLAB} (see Fig. 2) and the nucleon light-cone momentum distribution in iron (see Fig. 3) has been extracted in a  model-independent way.


\begin{thebibliography}{99}

\bibitem{Strikman} (a) See for a review L.L. Frankfurt and M.I. Strikman: 
 Phys. Rept. {\bf 160}, 235 (1988). (b) L.L. Frankfurt et al.: Phys. Rev. 
 {\bf C48}, 2451 (1993).

\bibitem{West} G.B. West: Phys. Rept. {\bf 18}, 263 (1975).

\bibitem{SLAC} D.B. Day et al.: Phys. Rev. Lett. {\bf 59}, 427 (1987).

\bibitem{FSI} C. Ciofi degli Atti and S. Simula: Phys. Lett. {\bf B325},  
 276 (1994) and references therein quoted.

\bibitem{Ciofi} C. Ciofi degli Atti, E. Pace and G.  Salm\'e: Phys. Rev. 
 {\bf C43}, 1155 (1991).

\bibitem{PKE} C. Ciofi degli Atti et al.: Phys. Rev. {\bf C44}, R7 (1991); 
 Phys. Rev. {\bf C53}, 1689 (1996).

\bibitem{note} A new scaling variable has been proposed in C. Ciofi 
 degli Atti and G.B. West: Phys. Lett. {\bf B458}, 447 (1999). However, its 
 definition relies on the (non-relativistic) model of \cite{PKE} and thus 
 binding effects are still treated in a model-dependent way.

\bibitem{EMC} U. Oelfke and P.U. Sauer: Nucl. Phys. {\bf A518}, 593 
 (1990). 

\bibitem{JLAB} J. Arrington et al.: Phys. Rev. Lett. {\bf 82}, 2056 (1999).

\bibitem{parton} R.P. Feynman: in {\em Photon-Hadron Interactions},
 Benjamin (New York, 1972). J.D. Bjorken and E.A. Paschos: Phys. Rev. {\bf 
 158}, 1975 (1969).

\bibitem{Wilson} K.G. Wilson: Phys. Rev. {\bf 179}, 1499 (1969).

\bibitem{GP76} H. Georgi and H.D. Politzer: Phys. Rev. {\bf D14}, 1829 
 (1976).

\bibitem{NAC73} O. Nachtmann: Nucl. Phys. {\bf B63}, 237 (1973).

\bibitem{Filippone} B.W. Filippone et al.: Phys. Rev. {\bf C45}, 1582 
 (1992); J. Arrington et al.: Phys. Rev. {\bf C53}, 2248 (1996).

\bibitem{Barbieri} R. Barbieri et al.: Phys. Lett. {\bf B64}, 171 (1976); 
 Nucl. Phys. {\bf B117}, 50 (1976); Phys. Lett. {\bf B81}, 207 (1979).

\bibitem{Ji} X. Ji and B.W. Filippone: Phys. Rev. {\bf C42}, R2279 (1990).

\bibitem{Wandzura} S. Wandzura: Nucl. Phys. {\bf B122}, 412 (1977).

\bibitem{SIM00} S. Simula:  Phys. Lett. {\bf B481}, 14 (2000).

\bibitem{SIM95} S. Simula: Few-Body Syst. Suppl. {\bf 8}, 423 (1995).

\end{thebibliography}
\end{document}